\documentstyle[prl,preprint,aps,epsf]{revtex}
\begin{document}

\title{Winding number transitions at finite temperature in the d = 2
Abelian-Higgs
       model}

\author{D.K. Park$^{1,2}$\footnote{e-mail:
dkpark@hep.kyungnam.ac.kr},
H. J. W. M\"{u}ller-Kirsten$^{1}$\footnote{e-mail:
mueller1@physik.uni-kl.de}and J.--Q.
Liang$^{1,3}$\footnote{e-mail:jqliang@mail.sxu.edu.cn}}
\address{1. Department of Physics, University of Kaiserslautern, D-67653 Kaiserslautern, Germany\\
2.Department of Physics, Kyungnam University, Masan, 631-701, Korea\\
3. Department of Physics and Institute of Theoretical Physics,
Shanxi University, Taiyuan, Shanxi 030006, China}
\maketitle

\maketitle
\begin{abstract}
Following our earlier investigations
we examine the quantum--classical winding
number transition in the Abelian-Higgs system. It is demonstrated that
the winding number transition in this system is of
the smooth second order type in the 
full range of parameter space. Comparison of the action of classical vortices
with that of the sphaleron supports our finding.
\end{abstract}

\vspace{3cm}

Recently much attention has been paid to the decay-rate transition between
the low-temperature instanton-dominated quantum tunneling regime and 
the high-temperature sphaleron-dominated thermal activity regime in
quantum mechanics\cite{chud97,liang98}, in field theoretical and gauge 
models\cite{hab96,park99-1,fro99,bon99}, 
and in cosmology\cite{gar94,fer95,kim99}. 
In particular the winding number transitions in gauge
theories are too complicated to handle analytically, and hence most
calculations of this type rely on numerical
simulation with the help of computers.
It is, however, usually difficult to obtain a good
physical insight from numerical 
calculations alone. Hence, it is important to develop
alternative methods which enable one
to extend the analytical approach as far as possible. 
Investigations along these directions were developed recently by using 
nonlinear perturbation theory\cite{gor97} or by counting
the number of negative
modes of the full Hessian around the sphaleron configuration\cite{lee99}.
Although these
two methods start from completely different points of view, they
both yield the
same criterion for a sharp first-order transition in the scalar 
field theories. 
Since the explicit form of the criterion is model-dependent, it is 
better to explain briefly how the criterion is derived at this stage.
Let $u_0$ and $\epsilon_0$ be eigenfunction
and eigenvalue of the negative mode of the fluctuation operator $\hat{h}$
around sphaleron. 
Therefore, 
the sphaleron frequency $\Omega_{sph}$ is defined as
$\Omega_{sph} \equiv \sqrt{-\epsilon_0}$. Then the type of the transition
is determined by computing the nonlinear corrections to the frequency. Let,
for example, $\Omega$ be a frequency involving the nonlinear corrections. 
If $\Omega_{sph} - \Omega < 0$, 
the energy dependence of the period of the periodic instanton becomes a
nonmonotonic function. This is easily conjectured from the fact that 
the energy-dependence of the period exhibits the increasing and
decreasing behaviours near the sphaleron and vacuum instanton. 
From this conjecture and the relation $d S / d \tau = E$ where 
$S$, $\tau$, and $E$ are classical action, period, and energy respectively,
one can imagine that 
the temperature dependence of instanton
action consists of monotonically decreasing and increasing parts
when $\Omega_{sph} - \Omega < 0$\cite{chud92}, which results
in the discontinuity in the derivative of action with respect to temperature
and hence, generates the sharp first-order transition. This is the 
main idea of Ref.\cite{park99-1,gor97}.

Some
applications of this criterion to condensed matter 
physics\cite{gor97,gor98,mul99}, 
field theoretical but non-gauge models\cite{park99-1,park99-2}, 
and cosmology\cite{kim99,liang99}
verify that this is physically reasonable. However,
the usefulness of such a criterion in the case of
gauge theories is not clear without an
application in a specific model.
This is clearly desirable since the winding number transitions in 
gauge theories imply additional complications such as gauge fixing
procedure and it is 
important to understand the implication of
those in 
physical phenomena such as baryon- and lepton-number violating processes.

In order to obtain some insight into
such transitions at higher temperatures, the criterion is
here applied to the Abelian--Higgs model,
which may be the simplest model among the gauge theories which support
both vacuum instanton and sphaleron configurations.

We start with the Euclidean
action of the d = 2 Abelian-Higgs model:
\begin{equation}
\label{action}
S_E = \int d\tau dx 
      \left[ \frac{1}{4} F_{\mu \nu} F_{\mu \nu} + (D_{\mu} \phi)^{\ast}
             D_{\mu} \phi + \lambda [\mid \phi \mid^2 - \frac{v^2}{2}]^2
                                                            \right]
\end{equation}
where $F_{\mu \nu} = \partial_{\mu} A_{\nu} - \partial_{\nu} A_{\mu}$ and
$D_{\mu} = \partial_{\mu} - igA_{\mu}$.

It is well-known that action (\ref{action}) is mathematically equivalent
to Ginzburg-Landau theory\cite{ginz50} and supports a vortex 
solution\cite{niel73} as a zero temperature solution. 
The temperature dependence of the classical
action for the periodic solution in this model is calculated
in Ref.\cite{mat93} using some special numerical techniques.
The final numerical result of Ref.\cite{mat93} shows that the
winding number phase transition in this model is of the smooth second-order
type in the range of $1/4 < M_H / M_W < 4$, where $M_H = \sqrt{2 \lambda} v$
and $M_W = g v$. In this paper we will follow the method developed in 
Ref.\cite{park99-1} and show that the type of the transition does not 
change over the full range of parameter space, {\it i.e.} it is always 
of the smooth
second order type.

The static solutions for the action (\ref{action}) whose field
equations are 
\begin{eqnarray}
\label{motion}
\partial_{\mu} F_{\mu \nu}&=& ig 
\left[ \phi^{\ast} (D_{\nu} \phi) - (D_{\nu} \phi)^{\ast} \phi \right]
                                                          \\ \nonumber
D_{\mu} D_{\mu} \phi&=& 2 \lambda \phi (\mid \phi \mid^2 - \frac{v^2}{2}), 
\end{eqnarray}
can be easily obtained:  
\begin{eqnarray}
\label{sphaleron}
A_0^{sph}&=&A_1^{sph} = 0  \\  \nonumber
\phi_{sph}&=& \frac{v}{\sqrt{2}} \tanh \sqrt{\frac{\lambda}{2}} v x.
\end{eqnarray}
In order to prove that $(A_0^{sph}, A_1^{sph}, \phi_{sph})$ are genuine
sphaleron configurations in this model, we introduce a non-contractible
loop\cite{for84,klei99,man83}
\begin{eqnarray}
\label{rev1}
\bar{A_0}&=& \bar{A_1} = 0    \\   \nonumber
\bar{\phi}&=& e^{is} \left[\frac{v}{\sqrt{2}} \cos s + i h(x) \sin s \right]
\end{eqnarray}
where $s$ is a loop parameter defined in the region $ 0 \leq s \leq \pi$.
Note that $\bar{\phi}$ becomes a trivial vacuum at the end points of $s$.
In addition, the minimizing condition of energy
${\cal E} (\phi, A) = \int dx {\cal L}_{E}$, where ${\cal L}_{E}$ is 
Euclidean Lagrangian density in Eq.(\ref{action}), makes $h(x)$ to be
\begin{equation}
\label{rev2}
h(x) = \frac{v}{\sqrt{2}} \tanh \left( \sin s \sqrt{\frac{\lambda}{2}} v x
                                                             \right).
\end{equation} 
Hence, $h(x)$ coincides with $\phi_{sph}$ when $s = \pi / 2$. It is 
easy to show that the energy along the minimal energy loop has a maximum
at $s = \pi / 2$, which proves that $(A_0^{sph}, A_1^{sph}, \phi_{sph})$
are sphaleron configuration. 

Chern-Simons number at $\tau = \tau_0$ in this model is defined as
\begin{equation}
\label{rev3}
{\cal N}_{cs} = \frac{1}{\pi v^2} \int_{-\infty}^{\tau_0} d \tau
                \int_{-\infty}^{\infty} d x \partial_{\mu}
                \Omega_{\mu}
\end{equation}
where the generalized Chern-Simons current $\Omega_{\mu}$ is 
\begin{equation}
\label{rev4}
\Omega_{\mu} = \epsilon_{\mu \nu} \left[ i \phi^{\ast} D_{\nu} \phi
               - \frac{g^2 v^2}{2} A_{\nu}  \right].
\end{equation}
In fact, $\partial_{\mu} \Omega_{\mu}$ is a lower bound of ${\cal L}_E$ when
$g = \sqrt{2 \lambda}$. To compute ${\cal N}_{cs}$ along the loop, we 
treat the loop parameter as an Euclidean time-dependent quantity
$s = s (\tau)$ with $s(\tau = - \infty) = 0$, $s(\tau = \infty) = \pi$,
and $s(\tau = \tau_0) = s_0$. Then it is straightforward to show that 
${\cal N}_{cs}$ along the loop is 
\begin{equation}
\label{rev5}
{\cal N}_{cs} = \frac{s_0}{\pi} - \frac{\sin 2 s_0}{2 \pi}.
\end{equation}
Hence, the sphaleron configuration($s_0 = \pi / 2$) has half-integer
Chern-Simons number whereas the trivial vacuum($s_0 = 0, \pi$) has 
integer one, which 
allows us to interpret 
the sphaleron as a classical solution sitting at the top of the barrier
separating the topologically distinct vacua.
The classical action corresponding to that
of the sphaleron is easily shown to be
\begin{equation}
\label{spaction}
S_{sph} = \frac{E_{sph}}{T_{sph}}
\end{equation}
where $T_{sph}$, the inverse of the sphaleron period, is interpreted
as a temperature and
\begin{equation}
\label{barrier}
E_{sph} = \frac{2 \sqrt{2 \lambda}}{3} v^3
\end{equation}
which is interpreted as the barrier height. 
Since the sphaleron is a static solution, one may wonder how to 
define the sphaleron period or frequency. In fact, the sphaleron 
frequency is defined by using a periodic instanton solution 
$\phi_{PI}(\tau, x;E)$ which is a time-dependent solution of the
Euclidean field equation (\ref{motion}) in the full range of energy
$0 < E < E_{sph}$. Since it is well known that $\phi_{PI}(\tau, x; E=0)$
and $\phi_{PI}(\tau, x; E = E_{sph})$ coincide with vacuum instanton
and sphaleron respectively, we define the sphaleron frequency is 
frequency of $\lim_{E \rightarrow E_{sph}} \phi_{PI}(\tau, x;E)$.

In order to be able to examine the type of quantum-classical
transition we have to 
introduce the fluctuation fields around the sphaleron and expand field
equations up to the third order in these fields. If, however, one expands
Eq.(\ref{motion}) naively, one will realize that the fluctuation operators
are not diagonalized and, hence, the spectra of these operators are not
obtainable analytically. To solve this problem we fix a gauge as a $R_{\xi}$ 
gauge\cite{car90-1,car90-2} by adding as gauge fixing term
\begin{equation}
\label{fixing}
S_{gf} = \frac{1}{2 \xi} \int d\tau dx
\left[ \partial_{\mu} A_{\mu} + \frac{ig}{2} \xi (\phi^2 - \phi^{\ast 2})
                                                           \right]^2
\end{equation}
to the original action (\ref{action}).
Then, the field equations are slightly changed to
\begin{eqnarray}
\label{eqmotion}
\partial_{\mu} F_{\mu \nu} + \frac{1}{\xi}
\left[ \partial_{\mu} \partial_{\nu} A_{\mu} + ig\xi (\phi \partial_{\nu}
       \phi - \phi^{\ast} \partial_{\nu} \phi^{\ast}) \right]
&=& ig \left[\phi^{\ast} (D_{\nu} \phi) - (D_{\nu} \phi)^{\ast} \phi \right]
                                             \\  \nonumber
D_{\mu}D_{\mu} \phi + ig\phi^{\ast}
\left[\partial_{\mu}A_{\mu} + \frac{ig\xi}{2} (\phi^2 - \phi^{\ast 2})
                                                            \right]
&=& 2 \lambda \phi (\mid \phi \mid^2 - \frac{v^2}{2}).
\end{eqnarray}
It is easy to show that the sphaleron solution (\ref{sphaleron}) and the
corresponding action (\ref{spaction}) are not changed under the 
$R_{\xi}$ gauge.

We now introduce the fluctuation fields around the sphaleron as follows:
\begin{eqnarray}
\label{fluc}
A_0(\tau, x)&=& a_0(\tau, x)    \nonumber \\ 
A_1(\tau, x)&=& a_1(\tau, x)    \\   
\phi(\tau, x)&=& \frac{1}{\sqrt{2}}
\left[ v \tanh \sqrt{\frac{\lambda}{2}} v x + \eta_1(\tau, x) + 
       i \eta_2(\tau, x)    \right]        \nonumber
\end{eqnarray}
where $a_0$, $a_1$, $\eta_1$, and $\eta_2$ are real fields.
After introducing the new space-time variables
\begin{eqnarray}
\label{nsp}
z_0&\equiv& \sqrt{\frac{\lambda}{2}} v \tau 
                                      \\  \nonumber
z_1&\equiv& \sqrt{\frac{\lambda}{2}} v x,
\end{eqnarray}
dimensionless parameters
\begin{equation}
\label{ddp}
\theta\equiv \frac{2 M_W}{M_H} = \sqrt{\frac{2 g^2}{\lambda}},
\end{equation}
and, for convenience, a function of $\theta$
\begin{equation}
s_1\equiv \sqrt{\theta^2 + \frac{1}{4}} - \frac{1}{2},
\end{equation}
one can show 
that at $\xi = 1$ the field equation (\ref{eqmotion}) can be expanded as
\begin{equation}
\label{expand}
\hat{l} \left( \begin{array}{c}
               a_0 \\ \rho_+ \\ \rho_- \\ \eta_1
               \end{array}                      \right)
= \hat{h}
        \left(  \begin{array}{c}
                a_0 \\ \rho_+ \\ \rho_- \\ \eta_1
                \end{array}                      \right)
+       \left(  \begin{array}{c}
                G_2^{a_0} \\ G_2^{\rho_+} \\ G_2^{\rho_-} \\ G_2^{\eta_1}
                \end{array}                       \right)
+       \left(  \begin{array}{c}
                G_3^{a_0} \\ G_3^{\rho_+} \\ G_3^{\rho_-} \\ G_3^{\eta_1}
                \end{array}                       \right)
\end{equation}
where                                
\begin{equation}
\label{explain}
\hat{l}=\left( \begin{array}{clcr}
                 \frac{\partial^2}{\partial z_0^2} & 0 & 0 & 0  \\
                 0 & \frac{\partial^2}{\partial z_0^2} & 0 & 0  \\
                 0 & 0 & \frac{\partial^2}{\partial z_0^2} & 0  \\
                 0 & 0 & 0 & \frac{\partial^2}{\partial z_0^2}  \\
                 \end{array}   \right),             
\hspace{2.0cm}
\hat{h} = \left( \begin{array}{clcr}
                 \hat{h}_{a_0} & 0 & 0 & 0  \\
                 0 & \hat{h}_{\rho_+} & 0 & 0  \\
                 0 & 0 & \hat{h}_{\rho_-} & 0  \\
                 0 & 0 & 0 & \hat{h}_{\eta_1}  \\
                 \end{array}                     \right), 
\end{equation}
and the functions $G_2$ and $G_3$ are given in the appendix 
(\ref{a1}).
Here, $\rho_+$ and $\rho_-$ are defined as
\begin{eqnarray}
\label{rohdef}
\rho_+&\equiv& \frac{1}{\sqrt{\cosh \alpha}}
\left[\cosh \frac{\alpha}{2} a_1 + \sinh \frac{\alpha}{2} \eta_2
                                                        \right], 
                                                             \\  \nonumber
\rho_-&\equiv& \frac{1}{\sqrt{\cosh \alpha}}
\left[-\sinh \frac{\alpha}{2} a_1 + \cosh \frac{\alpha}{2} \eta_2
                                                         \right]
\end{eqnarray}
where $\alpha = \sinh^{-1} 2 \theta$ and
\begin{eqnarray}
\label{defh}
\hat{h}_{a_0}&=& - \frac{\partial^2}{\partial z_1^2}
                 - \theta^2 {\rm sech}^2 z_1 + \theta^2,
                                                    \nonumber  \\
\hat{h}_{\rho_+}&=& - \frac{\partial^2}{\partial z_1^2}
                   - (s_1 - 1) s_1 {\rm sech}^2 z_1 + \theta^2,
                                                   \nonumber  \\
\hat{h}_{\rho_-}&=& - \frac{\partial^2}{\partial z_1^2}
                  - (s_1 + 1) (s_1 + 2) {\rm sech}^2 z_1 + \theta^2,
                                                            \\
\hat{h}_{\eta_1}&=& -\frac{\partial^2}{\partial z_1^2}
                    - 6 {\rm sech}^2 z_1 + 4.
                                                       \nonumber
\end{eqnarray}
The spatial parts of the fluctuation operators 
$\hat{h}_{a_0}$, $\hat{h}_{\rho_+}$, $\hat{h}_{\rho_-}$, and $\hat{h}_{\eta_1}$
are various kinds of P\"{o}schl-Teller type operators
whose spectra are 
summarized in Ref.\cite{jac77}. It is easy to show that the spectra of
$\hat{h}_{a_0}$ and $\hat{h}_{\rho_+}$ consist of only positive modes whose
explicit forms are not necessary for further study. What we need are only
the negative mode of $\hat{h}_{\rho_-}$ whose eigenfunction 
$\psi_{-1}^{(\rho_-)}$ and eigenvalue $\lambda_{-1}^{(\rho_-)}$ are
\begin{eqnarray}
\label{negamode}
\psi_{-1}^{(\rho_-)}(z_1)&=& 2^{-(s_1 + 1)}
\sqrt{\frac{\Gamma(2s_1 + 3)}{\Gamma(s_1 + 1) \Gamma(s_1 + 2)}}
\frac{1}{\cosh^{s_1 + 1} z_1},
                                                   \\   \nonumber
\lambda_{-1}^{(\rho_-)}&=& - s_1 - 1,
\end{eqnarray}
and the full spectrum of $\hat{h}_{\eta_1}$, which is summarized in Table I.
It is easy to show that the zero mode $\psi_{0}^{(\eta_1)}$ in 
Table I is propotional
to $\partial \phi_{sph} / \partial z_1$, which indicates the translational
symmetry of the Abelian-Higgs system.

Now, we have to carry out the perturbation to derive the criterion 
for the sharp first--order transition as suggested 
in Ref.\cite{gor97,mul99}. Since $\hat{l}$ and $\hat{h}$ in 
Eq. (\ref{expand}) are expressed in a matrix form, it is impossible
to use the criterion derived in Ref.\cite{gor97,mul99} directly.
In this case
we have to repeat the perturbation procedure with a spectrum of the full
spatial fluctuation operator $\hat{h}$ as suggested in Ref.\cite{park99-1}.
Computing the nonlinear corrections of the sphaleron
frequency $\Omega$ perturbatively, one can
derive the final result of the criterion for the sharp first-order
transition in this model as a following inequality:  
\begin{equation}
\label{criterion}
I_1(\theta, v) + I_2(\theta, v) + I_3(\theta, v)   <  0
\end{equation}
where 
\begin{eqnarray}
\label{iexplain}
I_1(\theta, v)&=&<\psi_{-1}^{(\rho_-)}(z_1) \mid D_1^{(1)} >,
                                                    \nonumber  \\
I_2(\theta, v)&=&<\psi_{-1}^{(\rho_-)}(z_1) \mid D_1^{(2)} >,
                                                               \\
I_3(\theta, v)&=&<\psi_{-1}^{(\rho_-)}(z_1) \mid D_1^{(3)} >.
                                                     \nonumber
\end{eqnarray}
Here $D_1^{(1)}(z_1)$,  $D_1^{(2)}(z_1)$, and $D_1^{(3)}(z_1)$ are given 
in the appendix (\ref{ddef}).
Since $T_{sph}$ in Eq. (\ref{spaction}) is 
the inverse of the sphaleron period, the action of the sphaleron becomes
\begin{equation}
\label{sphaction}
S_{sph} = \frac{8 \pi}{3 \sqrt{s_1 + 1}} v^2.
\end{equation}
In deriving $S_{sph}$ in Eq. (\ref{sphaction}) one has to use the 
rescaling definition of space-time variables (\ref{nsp}) and
$\Omega_{sph} = \sqrt{s_1 + 1}$ which is given in the appendix.

Now, in order to compute $D_1^{(i)}(z_1) \hspace{.5cm} i = 1, 2, 3$ 
we are in a position to compute $g_{\eta_1, 1}(z_1)$ and 
$g_{\eta_1, 2}(z_1)$ explicitly which is given in the appendix (\ref{gexpl}). 
The function $g_{\eta_1, 1}(z_1)$ is
explicitly derived as follows. We define
\begin{eqnarray*}
q_1(z_1) \equiv \hat{h}_{\eta_1}^{-1}
                 \frac{\sinh z_1}{\cosh^{2s_1 + 3} z_1}
\end{eqnarray*}
or equivalently
\begin{equation}
\label{defq}
\hat{h}_{\eta_1} q_1(z_1) = \frac{\sinh z_1}{\cosh^{2s_1 + 3} z_1}.
\end{equation}
Multiplying the zero mode of $\hat{h}_{\eta_1}$ with Eq. (\ref{defq}), 
integrating over $z_1$ from $-\infty$ to $z_1$, and performing partial
integration twice, we can obtain the first-order differential equation
for $q_1(z_1)$. Solving this differential equation one can obtain
$q_1(z_1)$ up to the constant of integration.
This constant is determined by the 
fact that $q_1(z_1)$ does not have a zero mode component. Inserting 
$q_1(z_1)$ into $g_{\eta_1, 1}(z_1)$ one can derive the explicit 
form of $g_{\eta_1, 1}(z_1)$ which is
\begin{equation}
\label{g11}
g_{\eta_1, 1}(z_1) = \frac{1}{4 \sqrt{\pi} v}
                    \frac{(s_1 - \frac{1}{2}) (s_1 + 1) 
                                        \Gamma(s_1 + \frac{1}{2})}
                                      {\Gamma(s_1 + 1)}
                    \frac{u(z_1)}{\cosh^2 z_1}
\end{equation}
where
\begin{equation}
\label{udeff}
u(z_1) = \int_0^{z_1} \frac{d y}{\cosh^{2 s_1} y}.
\end{equation}

Next we define
\begin{equation}
\label{defq2}
q_2(z_1) \equiv 
\left( \hat{h}_{\eta_1} + 4 \Omega_{sph}^2 \right)^{-1}
\frac{\sinh z_1}{\cosh^{2 s_1 +3} z_1}.
\end{equation}
Using the completeness condition as follows
\begin{equation}
\label{complete}
q_2(z_1) = \left( \hat{h}_{\eta_1} + 4 \Omega_{sph}^2 \right)^{-1}
          \left[
                \sum_{n=1}^{2}
                \mid \psi_n^{(\eta_1)} >
                < \psi_n^{(\eta_1)} \mid
                + \int dk 
                \mid \psi_k^{(\eta_1)}>
                <\psi_k^{(\eta_1)} \mid
                                           \right]
           \frac{\sinh z_1}{\cosh^{2 s_1 +3} z_1},
\end{equation}
one can obtain the integral representation of $q_2(z_1)$.
Inserting this into $g_{\eta_1, 2}(z_1)$, we can derive the final form
of $g_{\eta_1, 2}(z_1)$
\begin{eqnarray}
\label{g12}
g_{\eta_1, 2}(z_1)&=& \frac{1}{2 \sqrt{\pi} v}
                      \frac{(s_1 - \frac{1}{2}) (s_1 + 1) (s_1 + 2)
                            \Gamma(s_1 + \frac{1}{2})}
                           {\Gamma(s_1 + 1)}
                                                  \nonumber  \\
&\times& 
                    \Bigg[
                          \frac{3 \sqrt{\pi}}{4(4 s_1 + 7)}
                          \frac{\Gamma(s_1 + \frac{3}{2})}
                               {\Gamma(s_1 + 3)}
                          \frac{\sinh z_1}{\cosh^2 z_1}
                          + 
\frac{2^{2 s_1 + 2} (2 s_1 + 1) (2 s_1 + 3)}
                                 {2 \pi \Gamma(2 s_1 + 5)}
                                                               \\
& &
\hspace{1.0cm}
\times
                          \left[J_1(\theta, z_1) + 3 \tanh z_1 
                                 \left( J_2(\theta, z_1)
                                             - J_4(\theta, z_1) \right)
                                 - 3 \tanh^2 z_1 J_3(\theta, z_1)
                                                           \right]
                                                             \Bigg]
                                                     \nonumber
\end{eqnarray}     
where
\begin{eqnarray}
\label{jjdef}
J_1(\theta, z_1)&\equiv&
            \int_0^{\infty} dk
            \frac{k \Gamma(s_1 + 1 + \frac{i k}{2} )
                     \Gamma(s_1 + 1 - \frac{i k}{2} )}
                 {4 (s_1 + 2) + k^2}
                   \sin k z_1,
                                    \nonumber  \\
J_2(\theta, z_1)&\equiv&
            \int_0^{\infty} dk
            \frac{ \Gamma(s_1 + 1 + \frac{i k}{2} )
                     \Gamma(s_1 + 1 - \frac{i k}{2} )}
                 {4 (s_1 + 2) + k^2}
                   \cos k z_1,
                                       \nonumber  \\
J_3(\theta, z_1)&\equiv&
            \int_0^{\infty} dk
            \frac{k \Gamma(s_1 + 1 + \frac{i k}{2} )
                     \Gamma(s_1 + 1 - \frac{i k}{2} )}
                 {(1 + k^2)[4 (s_1 + 2) + k^2]}
                   \sin k z_1,                                                              
                                                  \\
J_4(\theta, z_1)&\equiv&
            \int_0^{\infty} dk
            \frac{ \Gamma(s_1 + 1 + \frac{i k}{2} )
                     \Gamma(s_1 + 1 - \frac{i k}{2} )}
                 {(1 + k^2)[4 (s_1 + 2) + k^2]}
                   \cos k z_1.
                                               \nonumber
\end{eqnarray}
                                                    
Now the computation of $I_1(\theta, v)$, $I_2(\theta, v)$, and 
$I_3(\theta, v)$ is straightforward and their final form is given in 
the appendix (\ref{final}).
It is very interesting that $I_1(\theta, v)$ and $I_2(\theta, v)$ vanish
at $s_1 = 1 / 2$ or, in terms of $\theta$, at $\theta = \sqrt{3} / 2$.

The $\theta$-dependence of $I_1(\theta, v)$, $I_2(\theta, v)$,
$I_3(\theta, v)$, and $I_1(\theta, v) + I_2(\theta, v) + I_3(\theta, v)$
is shown in Fig. 1 which shows that the condition for the sharp 
first-order transition, i.e.  (\ref{criterion}),
does not hold 
when $\theta < 4$. Ploting $I_1(\theta, v) + I_2(\theta, v) + I_3(\theta, v)$
in the range of large $\theta$, one can confirm numerically
that it is a monotonically increasing function which indicates that 
the sharp first-order transition does not occur in the full range of 
parameter space.
This means the winding number phase transition of this model is always
smooth second-order as shown in Ref.\cite{mat93}, where same conclusion
was derived 
by numerical method in the restricted region of parameter
space.

There is another indirect method
which confirms our conclusion.
If the transition is second order and 
there is no interaction between vacuum instanton and anti-instanton,
the condition
\begin{equation}
\label{cond1}
2 S_1 > S_{sph},
\end{equation}
where $S_1$ is the action of one instanton solution, has to be satisfied.
Since there is no interaction between vortices at the Bogomol'nyi 
limit\cite{bog76,vega76} which is $\theta = 2$ in this model, we can use 
the condition (\ref{cond1}) to check the credibility of our conjecture.   
Since $S_1 = \pi v^2$ in this limit, it is easy to show that 
\begin{equation}
\label{ratio}
\frac{S_{sph}}{2 S_1} = 0.833 < 1
\end{equation}
which supports our conclusion.

In general, there is an interaction between vortices and hence, the 
condition (\ref{cond1}) has to be modified to
\begin{equation}
\label{cond2}
S_2 > S_{sph},
\end{equation}
where $S_2$ is action of two interacting vortices, at arbitrary $\theta$.
$S_1$ and $S_2$ at arbitrary $\theta$ 
can be computed numerically by employing the variation 
method\cite{jaco79}. $S_1$, $S_2$ and $S_{sph}$ at various $\theta$ are 
summarized at Table II, which also confirms our finding at 
$0 < \theta < 4$.

We hope our method can be applicable to the $SU(2)$-Higgs model
which is most important to understand the baryon-number violating
process. The approach along this direction is under investigation.

\vskip 1cm
{\bf Acknowledgement:} D.K. P. acknowledges support by DAAD and the
Deutsche Forschungsgemeinschaft(DFG), J.--Q. L. acknowledges 
support
by a DAAD--K.C. Wong Fellowship and the Deutsche Forschungsgemeinschaft(DFG).
This work was also supported by Korea Research Foundation(1999-015-DP0074).

\begin{table}
\begin{tabular}{|c|c|}
 Eigenvalue of $\hat{h}_{\eta_1}$ &  Eigenfunction of $\hat{h}_{\eta_1}$
                                              \hspace{2.0cm}   \\
                                        \hline 
 $\lambda_0^{(\eta_1)} = 0$ & $\psi_0^{(\eta_1)}(z_1) = \frac{\sqrt{3}}{2}
                                                  \frac{1}{\cosh^2 z_1}$  
                                                \hspace{2.0cm}  \\
                                        \hline 
 $\lambda_1^{(\eta_1)} = 3$ & $\psi_1^{(\eta_1)}(z_1) = \sqrt{\frac{3}{2}}
                                                  \frac{\sinh z_1}
                                                       {\cosh^2 z_1}$   
                                                  \hspace{2.0cm} \\
                                        \hline 
 $\lambda_k^{(\eta_1)} = 4 + k^2$ & $\psi_k^{(\eta_1)}(z_1) = 
                                      - \frac{1}{\sqrt{2\pi}}
                                       \frac{e^{i k z_1}}
                                            {(1 + i k) (2 + i k)}
                                     \left[ (1 + k^2) + 3 i k \tanh z_1
                                            - 3 \tanh^2 z_1
                                                             \right]$  
                                             \hspace{2.0cm}  \\                            
\end{tabular}
\vspace{0.5cm}
\caption{Eigenvalues and eigenfunctions of $\hat{h}_{\eta_1}$}
\end{table}

\begin{table}
\begin{tabular}{|c|c|c|c|}
$\theta$ & $S_{sph} / \pi v^2$ \hspace{2.0cm} & $S_1 / \pi v^2$ \hspace{2.0cm} & $S_2 / \pi v^2$   \hspace{2.0cm} \\
                                               \hline
$0.5$   & $2.43$     \hspace{2.0cm} & $1.79$ \hspace{2.0cm} & $4.31$  \hspace{2.0cm} \\
                                               \hline
$1.0$   & $2.10$     \hspace{2.0cm} & $1.34$ \hspace{2.0cm} & $2.94$  \hspace{2.0cm} \\
                                               \hline
$1.5$   & $1.85$     \hspace{2.0cm} & $1.13$ \hspace{2.0cm} & $2.34$  \hspace{2.0cm} \\
                                               \hline
$2.0$   & $1.67$     \hspace{2.0cm} & $1.00$ \hspace{2.0cm} & $2.00$  \hspace{2.0cm} \\
                                               \hline
$2.5$   & $1.53$     \hspace{2.0cm} & $0.91$ \hspace{2.0cm} & $1.78$  \hspace{2.0cm} \\ 
                                               \hline
$3.0$   & $1.42$     \hspace{2.0cm} & $0.85$ \hspace{2.0cm} & $1.62$  \hspace{2.0cm} \\
                                               \hline
$3.5$   & $1.33$     \hspace{2.0cm} & $0.80$ \hspace{2.0cm} & $1.50$  \hspace{2.0cm} \\
                                               \hline
$4.0$   & $1.25$     \hspace{2.0cm} & $0.76$ \hspace{2.0cm} & $1.40$  \hspace{2.0cm} \\
\end{tabular}
\vspace{0.5cm}
\caption{$S_1$, $S_2$ and $S_{sph}$ at various values of $\theta$}
\end{table}
\begin{appendix}{\centerline{\bf Appendix A}}

\setcounter{equation}{0}
\renewcommand{\theequation}{A.\arabic{equation}}
In this appendix we collect the lengthy expressions to make the 
main text to be simple and compact. 

In the expansion of equation of motion (\ref{expand}) the higher order
terms $G_2$ and $G_3$ are
\begin{eqnarray}
\label{a1}
G_2^{a_0}&=& \frac{1}{v}
\left[ \theta \sqrt{\frac{2 s_1}{s_1 + \frac{1}{2}}} \rho_+
       \frac{\partial \eta_1}{\partial z_0}
       + \theta \sqrt{\frac{2 (s_1 + 1)}{s_1 + \frac{1}{2}}} \rho_-
       \frac{\partial \eta_1}{\partial z_0}
       + 2 \theta^2 \tanh z_1 a_0 \eta_1           \right], 
                                                            \nonumber \\
G_3^{a_0}&=& \frac{1}{v^2}
\left[ \theta^2 a_0 \eta_1^2 + \frac{\theta^2 s_1}{2(s_1 + \frac{1}{2})}
       a_0 \rho_+^2 + \frac{\theta^2 (s_1 + 1)}{2 (s_1 + \frac{1}{2})}
       a_0 \rho_-^2 + \frac{\theta^3}{s_1 + \frac{1}{2}} a_0 \rho_+ \rho_-
                                                              \right],
                                                            \nonumber \\
G_2^{\rho_+}&=& \frac{1}{v}
\Bigg[ \theta \sqrt{\frac{2 s_1}{s_1 + \frac{1}{2}}} a_0 
       \frac{\partial \eta_1}{\partial z_0}
     + \frac{2 \theta^2}{s_1 + \frac{1}{2}} \rho_+ 
       \frac{\partial \eta_1}{\partial z_1}
     + \frac{\theta}{s_1 + \frac{1}{2}} \rho_- 
       \frac{\partial \eta_1}{\partial z_1}      
                                                      \nonumber  \\
     & & 
\hspace{2.0cm}
      + \tanh z_1
      \left[ 2 \left( \theta^2 + \frac{s_1}{s_1 + \frac{1}{2}} \right)
               \rho_+ \eta_1 + \frac{2 \theta}{s_1 + \frac{1}{2}}
               \rho_- \eta_1        
                                                      \right]
                                                             \Bigg],
                                                            \nonumber \\
G_3^{\rho_+}&=& \frac{1}{v^2}
   \Bigg[
         \frac{\theta^2 s_1}{2(s_1 + \frac{1}{2})} a_0^2 \rho_+
        +\frac{\theta^3}{2(s_1 + \frac{1}{2})} a_0^2 \rho_-
        + \left( \theta^2 + \frac{s_1}{s_1 + \frac{1}{2}} \right)
                                     \rho_+ \eta_1^2
        +\frac{\theta}{s_1 + \frac{1}{2}} \rho_- \eta_1^2
        +\frac{1}{2} \frac{s_1^2 + \theta^4}{(s_1 + \frac{1}{2})^2}
                                                    \rho_+^3
                                                           \nonumber \\
& & 
\hspace{1.0cm}
        +\frac{3 \theta}{2(s_1 + \frac{1}{2})^2} 
                  \left( s_1 + \frac{\theta^2}{2} \right) \rho_+^2 \rho_-
        +\frac{\theta}{2(s_1 + \frac{1}{2})^2}
                  \left( s_1 + 1 - \frac{\theta^2}{2} \right) \rho_-^3
        +\frac{\theta^2 (7 - 2 \theta^2)}{4(s_1 + \frac{1}{2})^2}
                                 \rho_+ \rho_-^2
                                                       \Bigg],
                                                         \nonumber \\
G_2^{\rho_-}&=& \frac{1}{v}
\Bigg[ \theta \sqrt{\frac{2(s_1 + 1)}{s_1 + \frac{1}{2}}}
               a_0 \frac{\partial \eta_1}{\partial z_0}
      - \frac{2 \theta^2}{s_1 + \frac{1}{2}} \rho_-
        \frac{\partial \eta_1}{\partial z_1}
      + \frac{\theta}{s_1 + \frac{1}{2}} \rho_+ 
        \frac{\partial \eta_1}{\partial z_1}  
                                                   \\  
& &
\hspace{2.0cm}
      + \tanh z_1
        \left[ 2 \left( \theta^2 + 
                        \frac{s_1 + 1}{s_1 + \frac{1}{2}}  \right)
                 \rho_- \eta_1 + \frac{2 \theta}{s_1 + \frac{1}{2}}
                                 \rho_+ \eta_1
                                                 \right]
                                                                \Bigg],
                                                         \nonumber \\
G_3^{\rho_-}&=&\frac{1}{v^2}
\Bigg[
      \frac{\theta^2(s_1 + 1)}{2(s_1 + \frac{1}{2})} a_0^2 \rho_-
     +\frac{\theta^3}{2(s_1 + \frac{1}{2})} a_0^2 \rho_+
     + \left( \theta^2 + \frac{s_1 + 1}{s_1 + \frac{1}{2}} \right)
             \rho_- \eta_1^2
     + \frac{\theta}{s_1 + \frac{1}{2}} \rho_+ \eta_1^2
                                                        \nonumber \\
& &
\hspace{.7cm}
     + \frac{\theta^4 + (s_1 + 1)^2}{2(s_1 + \frac{1}{2})^2} \rho_-^3
     + \frac{\theta (s_1 + \frac{\theta^2}{2})}{2(s_1 + \frac{1}{2})^2}
                                      \rho_+^3
     +\frac{3 \theta (s_1 + 1 - \frac{\theta^2}{2})}{2(s_1 + \frac{1}{2})^2}
                           \rho_+ \rho_-^2
     +\frac{\theta^2 (7 - 2 \theta^2)}{4(s_1 + \frac{1}{2})^2}
                           \rho_+^2 \rho_-
                                                            \Bigg],
                                                              \nonumber \\
G_2^{\eta_1}&=& \frac{1}{v}
\Bigg[
      - \theta \sqrt{\frac{2 s_1}{s_1 + \frac{1}{2}}}
        \left( \frac{\partial a_0}{\partial z_0} \rho_+
              +a_0 \frac{\partial \rho_+}{\partial z_0} 
                                                       \right)
      - \theta \sqrt{\frac{2(s_1 + 1)}{s_1 + \frac{1}{2}}}
        \left( \frac{\partial a_0}{\partial z_0} \rho_-
              +a_0 \frac{\partial \rho_-}{\partial z_0}  
                                                       \right)
                                                              \nonumber \\
& & 
\hspace{.7cm}
      -\frac{2 \theta^2}{s_1 + \frac{1}{2}}
       \left( \rho_+ \frac{\partial \rho_+}{\partial z_1}
             - \rho_- \frac{\partial \rho_-}{\partial z_1} 
                                                        \right)
      -\frac{\theta}{s_1 + \frac{1}{2}}
        \left( \frac{\partial \rho_+}{\partial z_1} \rho_- 
              + \rho_+ \frac{\partial \rho_-}{\partial z_1}
                                                        \right)
                                                            \nonumber  \\
& &
\hspace{.7cm}
      +\tanh z_1
       \left[ \theta^2 a_0^2 + 6 \eta_1^2
             + \left( \theta^2 + \frac{s_1}{s_1 + \frac{1}{2}} \right)
               \rho_+^2
             + \left( \theta^2 + \frac{s_1 + 1}{s_1 + \frac{1}{2}} \right)
               \rho_-^2
             + \frac{2 \theta}{s_1 + \frac{1}{2}} \rho_+ \rho_-
                                                               \right]
                                                                 \Bigg],
                                                            \nonumber \\
G_3^{\eta_1}&=& \frac{1}{v^2}
\left[
      \theta^2 a_0^2 \eta_1 + 
      \left(\theta^2 + \frac{s_1}{s_1 + \frac{1}{2}} \right) \rho_+^2 \eta_1
     + \left( \theta^2 + \frac{s_1 + 1}{s_1 + \frac{1}{2}} \right)
                                                       \rho_-^2 \eta_1
     + \frac{2 \theta}{s_1 + \frac{1}{2}} \rho_+ \rho_- \eta_1
     + 2 \eta_1^3                              \right].
                                                         \nonumber 
\end{eqnarray}

In Eq. (\ref{iexplain}) $D_1^{i}(z_1) \hspace{.2cm} i = 1, 2, 3$ are
\begin{eqnarray}
\label{ddef}
D_1^{(1)}(z_1)&=& \frac{2^{-s_1}}{v}
                  \sqrt{\frac{\Gamma(2s_1 + 3)}{\Gamma(s_1 + 1) 
                                                 \Gamma(s_1 + 2)}} 
                                                        \nonumber \\
& &
\hspace{.7cm}
\times
\left[
      \frac{(s_1 + 1) (2 s_1^2 + s_1 + 2)}{2 s_1 + 1}
      \frac{\sinh z_1}{\cosh^{s_1 + 2} z_1} g_{\eta_1, 1}
    - \frac{s_1 (s_1 + 1)}{s_1 + \frac{1}{2}}
      \frac{1}{\cosh^{s_1 + 1} z_1}
      \frac{d g_{\eta_1, 1}}{d z_1}   \right],
                                                     \nonumber \\
D_1^{(2)}(z_1)&=& \frac{2^{-(s_1 + 1)}}{v}
                  \sqrt{\frac{\Gamma(2s_1 + 3)}{\Gamma(s_1 + 1) 
                                                 \Gamma(s_1 + 2)}}
                                                      \\
& &
\hspace{.7cm}
\times
\left[
      \frac{(s_1 + 1) (2 s_1^2 + s_1 + 2)}{2 s_1 + 1}
      \frac{\sinh z_1}{\cosh^{s_1 + 2} z_1} g_{\eta_1, 2}
    - \frac{s_1 (s_1 + 1)}{s_1 + \frac{1}{2}}
      \frac{1}{\cosh^{s_1 + 1} z_1}
      \frac{d g_{\eta_1, 2}}{d z_1}   \right], 
                                                  \nonumber  \\
D_1^{(3)}&=& \frac{3 \cdot 2^{-3 s_1 - 6}}{v^2}
             (1 + s_1^2)
             \left( \frac{s_1 + 1}{s_1 + \frac{1}{2}} \right)^2
             \left( \frac{\Gamma(2 s_1 + 3)}{\Gamma(s_1 + 1) \Gamma(s_1 + 2)}
                                                  \right)^{\frac{3}{2}}
             \frac{1}{\cosh^{3 s_1 + 3} z_1},   
                                                    \nonumber
\end{eqnarray}
where 
\begin{eqnarray}
\label{gexpl}
g_{\eta_1, 1}(z_1)&=& \frac{1}{2 \sqrt{\pi} v}
                      \frac{\Gamma(s_1 + \frac{1}{2})}{\Gamma(s_1 + 1)}
                      (s_1 - \frac{1}{2}) (s_1 + 1) (s_1 + 2)
                      \hat{h}_{\eta_1}^{-1}
                      \frac{\sinh z_1}{\cosh^{2 s_1 + 3} z_1},
                                                     \nonumber  \\
g_{\eta_1, 2}(z_1)&=& \frac{1}{2 \sqrt{\pi} v}
                      \frac{\Gamma(s_1 + \frac{1}{2})}{\Gamma(s_1 + 1)}
                      (s_1 - \frac{1}{2}) (s_1 + 1) (s_1 + 2)
                      \left( \hat{h}_{\eta_1} + 4 \Omega_{sph}^2 \right)^{-1}
                      \frac{\sinh z_1}{\cosh^{2 s_1 + 3} z_1},
\end{eqnarray}
and $\Omega_{sph} = \sqrt{-\lambda_{-1}^{(\rho_-)}} = \sqrt{s_1 + 1}$ is 
the zeroth order frequency of the sphaleron.

Using the explicit results of $g_{\eta_1, 1}(z_1)$ and 
$g_{\eta_1, 2}(z_1)$ it is straightforward to calculate
$I_1(\theta, v)$, $I_2(\theta, v)$, and $I_3(\theta, v)$ given in 
Eq. (\ref{iexplain}): 
\begin{eqnarray}
\label{final}
I_1(\theta, v)&=& - \frac{1}{4 \sqrt{\pi}v^2}
                    (s_1 - \frac{1}{2})^2 (s_1 + 1)^2
                  \frac{\Gamma^2(s_1 + \frac{1}{2})
                        \Gamma(2 s_1 + 2)}
                       {\Gamma^2(s_1 + 1)
                        \Gamma(2 s_1 + \frac{5}{2})},
                                             \nonumber   \\
I_3(\theta, v)&=& \frac{3 \cdot 2^{2 s_1 - 2}}{v^2 \pi}
                  (1 + s_1^2)
                  \left( \frac{s_1 + 1}{s_1 + \frac{1}{2}} \right)^2
                  \frac{\Gamma^3 (s_1 + \frac{3}{2})}
                       {\Gamma(s_1 + 1) \Gamma(2 s_1 + \frac{5}{2})},
                                                       \nonumber  \\
I_2(\theta, v)&=& - \frac{1}{2 \pi v^2}
                  \frac{(s_1 - \frac{1}{2}) (s_1 + 1) (s_1 + 2)}
                       {(s_1 + \frac{1}{2})}
                  \frac{\Gamma^2 (s_1 + \frac{3}{2})}
                       {\Gamma^2 (s_1 + 1)}     
                                                      \\
&\times&
                 \Bigg[
                       \frac{3 \pi (2 s_1 - 1) (s_1 + 1) (s_1 + 2)}
                            {4 (2 s_1 + 1) (4 s_1 + 7)}
                       \frac{\Gamma^2 (s_1 + \frac{3}{2})}
                            {\Gamma^2 (s_1 + 3)}
                      +
                                                 \nonumber \\
& &
\hspace{.6cm}
\frac{2^{2 s_1 + 1} (s_1 - 2) (2 s_1 + 1) (2 s_1 + 3)}
                            {\pi \Gamma(2 s_1 + 5)} 
                        \int_0^{\infty} d z_1
                            \frac{J_5(\theta, z_1)}
                                 {\cosh^{2 s_1 + 2} z_1}
                       -
                                                   \nonumber  \\
& &
\hspace{.6cm}
\frac{3 \cdot 2^{2 s_1 + 2} (2 s_1 + 3) 
                              (s_1 + \frac{1}{2}) (2 s_1^2 + 3 s_1 + 2)}
                             {\pi \Gamma(2 s_1 + 5)}
                         \int_0^{\infty} d z_1
                             \frac{J_2(\theta, z_1) - J_4(\theta, z_1)}
                                  {\cosh^{2 s_1 + 2} z_1}
                        +
                                                    \nonumber  \\
& &
\hspace{.6cm}
\frac{3 \cdot 2^{2 s_1 + 2} (s_1 + 1)
                               (s_1 + \frac{3}{2}) (2 s_1 + 1) (2 s_1 + 3)}
                              {\pi \Gamma(2 s_1 + 5)}
                         \int_0^{\infty} d z_1
                             \frac{J_2(\theta, z_1) - J_4(\theta, z_1)}
                                  {\cosh^{2 s_1 + 4} z_1}
                                                   \Bigg],
                                                         \nonumber
\end{eqnarray}
where 
\begin{equation}
J_5(\theta, z_1) \equiv
                 \int_0^{\infty}
                 \frac{k^2 \Gamma(s_1 + 1 + \frac{i k}{2})
                           \Gamma(s_1 + 1 - \frac{i k}{2})}
                      {4 (s_1 + 2) + k^2}
                      \cos k z_1.
\end{equation}

\end{appendix} 

\begin{figure}

\caption{The $\theta$-dependence of $I_1$, $I_2$, $I_3$, and $I_1 + I_2 + I_3$
         at $v = 1$. From this figure we can conclude that the winding
         number
         transition of the Abelian-Higgs model is smooth second-order in the
         full parameter range.}
\end{figure}

%
\newpage
\epsfysize=20cm \epsfbox{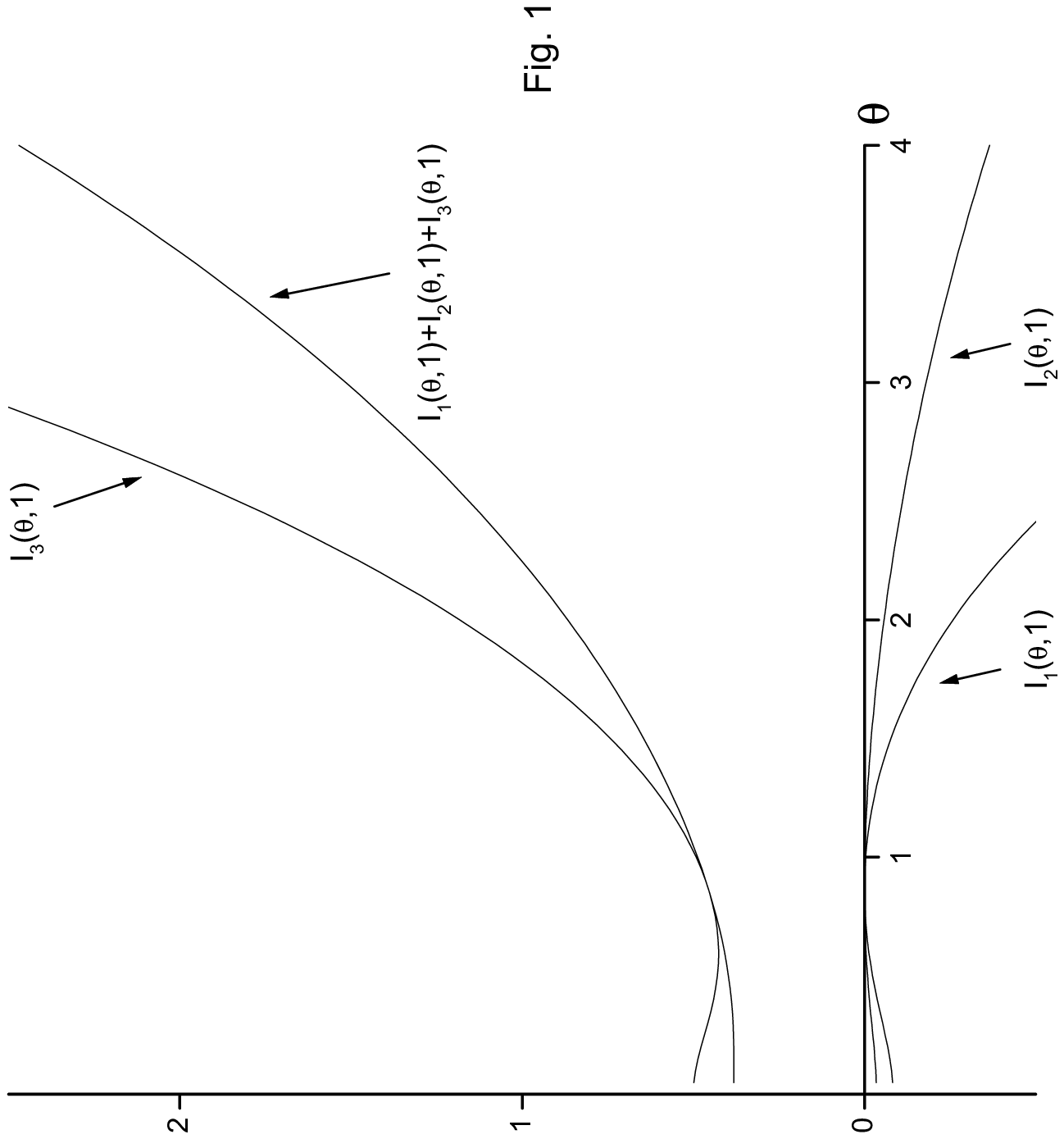}

\end{document}